# Ultrasensitive piezoelectric sensor based on two-dimensional Na₂Cl crystals with periodic atom vacancies

Tao Wang[1,#], Yan Fan[2,#], Jie Jiang[1,#], Yangyang Zhang[3,#], Yingying Huang[4,*], Liuyuan Zhu[4], Haifei Zhan[5], Chunli Zhang[6], Bingquan Peng[7], Zhen Gu[8], Qiubo Pan[4], Junjie Wu[6], Junlang Chen[2], Pei Li[1], Lei Zhang[9], Liang Chen[1,*], Chaofeng Lü[3,5,*], and Haiping Fang[4,10,*]

[1]*School of Physical Science and Technology, Ningbo University, Ningbo, 315211, China.*
[2]*Department of Optical Engineering, Zhejiang A&F University, Hangzhou, 311300, China.*
[3]*Faculty of Mechanical Engineering & Mechanics, Ningbo University, Ningbo, 315211, China.*
[4]*School of Physics, East China University of Science and Technology, Shanghai, 200237, China.*
[5]*College of Civil Engineering and Architecture, Key Laboratory of Soft Machines and Smart Devices of Zhejiang Province, Zhejiang University, Hangzhou 310027, China.*
[6]*Department of Engineering Mechanics, Zhejiang University, Hangzhou 310027, China.*
[7]*Wenzhou Institute, University of Chinese Academy of Sciences, Wenzhou, Zhejiang 325000, China.*
[8]*Key Laboratory of Smart Manufacturing in Energy Chemical Process Ministry of Education, East China University of Science and Technology, Shanghai 200237, China.*
[9]*MOE Key Laboratory for Nonequilibrium Synthesis and Modulation of Condensed Matter, School of Physics, Xi'an Jiaotong University, Xi'an 710049, China.*
[10]*School of Physics, Zhejiang University, Hangzhou 310027, China.*

[#]These authors contributed equally to this work.
*Corresponding author. E-mail: fanghaiping@sinap.ac.cn (H.F.); lucf@nbu.edu.cn (C.L.); liangchen@zafu.edu.cn (L.C.); and huangyingying@ecust.edu.cn (Y. H.)

**Pursuing ultrasensitivity of pressure sensors has been a long-standing goal. Here, we report a piezoelectric sensor that exhibits supreme pressure-sensing performance, including a peak sensitivity up to $3.5 \times 10^6$ kPa$^{-1}$ in the pressure range of 1-100 mPa and a detection limit of less than 1 mPa, superior to the current state-of-the-art pressure sensors. These properties are attributed to the high percentage of periodic atom vacancies in the two-dimensional Na₂Cl crystals formed within multilayered graphene oxide membrane in the sensor, which provides giant polarization with high stability. The sensor can even clearly detect the airflow fluctuations surrounding a flapping butterfly, which have long been the elusive tiny signals in the famous "butterfly effect". The finding represents a step towards next-generation pressure sensors for various precision applications.**

Pursuing pressure sensors that offer precise mechano-electric transduction of tiny mechanical stimuli under complex ambient conditions, with properties such as high sensitivity, low pressure detection limit and a broad pressure-response range, is of great importance and has been a long-standing goal (*1-9*). Microstructures, such as micro-structures nano-structures, and even atomic structures, exhibit inextricable link with the promotion of sensitivity and detection limit for pressure sensors due to their asymmetrical structures (*5, 6, 10-18*). Micro-structures can effectively improve sensitivity, including micro-pyramid arrays (*5, 6, 19-21*), wrinkles (*22*), micro-domes (*23, 24*), micro-pillar arrays (*10*), and micro-protrusions (*25*), as well as other assembly patterns (*11-14, 26-28*). More precisely, asymmetrical nano-structures and heterogeneities (*15-17, 29-32*) are recognized to play a key role in significantly promoting the mechano-electric transduction of various stimuli. Down to atom scale of atom vacancies, very large piezoelectric sensitivities have been observed by using extremely strong electric field (*18*). But under ambient conditions, numerous atom vacancies may make the structure difficult to maintain stable.

Here, we present a piezoelectric sensor with supreme pressure-sensing performance in the sensitivity and detection limit, based on a high percentage of atom vacancies in the unique two-dimensional (2D) $Na_2Cl$ crystals with an unconventional stoichiometry of Na to Cl atoms within multilayered graphene oxide (GO) membrane. The atom vacancies result in a very high piezoelectric coefficient of the membrane. The periodic arrangement of these atom vacancies ensures the structural stability. Remarkably, the sensor is capable of detecting signals as low as 1 mPa (corresponding to a 25 nanonewton force), together with excellent mechanical stability. The small size, fast response, and self-powered capability of the sensor make it particularly suitable for use in biological systems and complex environments. In addition, the application scope of the sensor can be greatly extended to encompass the detection of other tiny signals including electrical, magnetic, optical, and thermal signals, by converting these signals to mechanical forces.

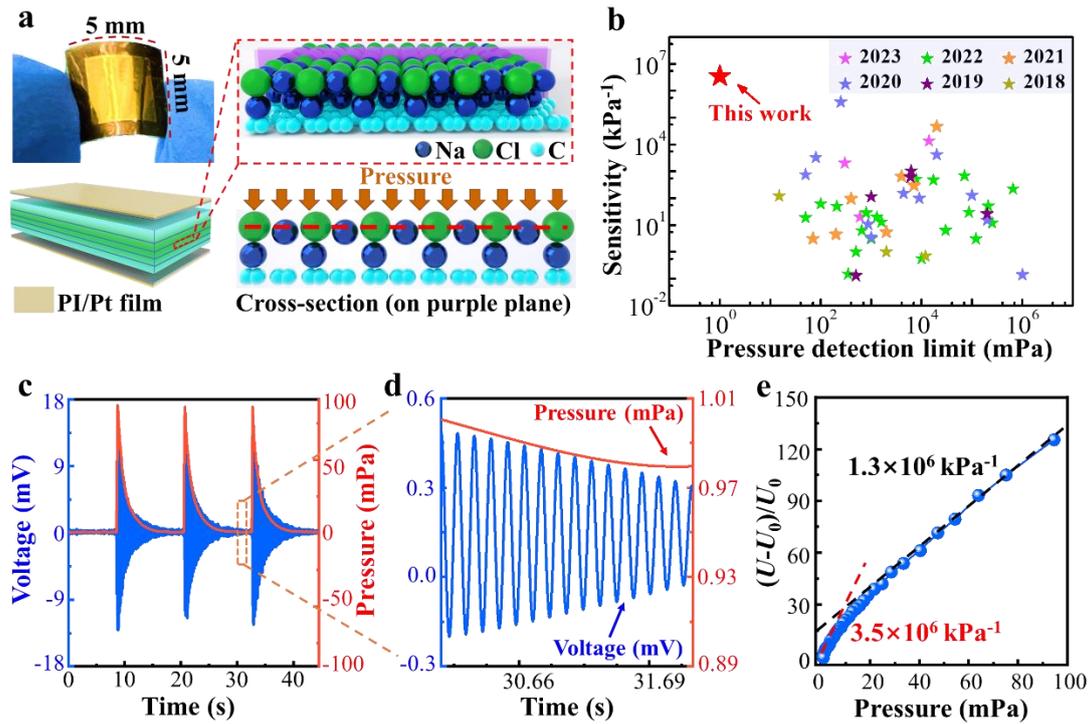

**Figure 1 | Piezoelectric sensor with supreme pressure-sensing performance. a**, Schematic of a flexible piezoelectric sensor based on $Na_2Cl$ crystals/GO membranes between conductive Pt/polyimide (PI) films. The enlarged views depict the microstructure of the $Na_2Cl$ crystal/GO membrane and show a cross-section of an atom-thick slice, as well as mechanical deformation under external pressure. **b**, Comparison of sensitivity and the pressure detection limit of the proposed sensor and other high-performance pressure sensors (for details, see Table S1). **c**, Decay process of the acoustic excitation with ~100 mPa and 260 Hz in a tuning fork test. A bandpass filter ranging from 250 to 270 Hz was applied to the output voltage to remove background signals, as well as 50 Hz alternating current (AC) signal and its harmonics. **d**, An enlarged view of the voltage response of the sensor. **e**, Normalized relative change in voltage $((U-U_0)/U_0)$ as a function of acoustic pressure below 100 mPa. $U_0 = 0.1$ mV is the initial voltage of the sensor without pressure, and $U$ is the voltage under the applied pressure $p$. The dashed red and black lines in the figure are the tangent lines on the curve, and their slopes are the values of sensitivity $S$.

The membranes with GO layers and 2D Na-Cl crystals were prepared by spray-coating GO suspension and dilute NaCl solution on the polyimide (PI) substrate for several times (Supplementary Information, Materials and Method). These membranes, having a thickness of ~5 μm, were removed from the substrate and cut into ~5 × 5 mm² sections. Then, they were placed between conductive Pt/PI substrates as sensor devices (Fig. 1a).

The sensor exhibits supreme sensitivity for pressure sensing in the low-pressure range of 1–100 mPa. An acoustic pressure below 100 mPa was generated with the acoustic excitation of a tuning fork with a frequency of ~260 Hz. The acoustic pressure ($P$) of the acoustic excitation was monitored using an acoustic tester (UMM-6, Dayton Audio).

The voltage response of the sensor (*U*) to changes in the acoustic pressure was simultaneously monitored by a digital lock-in amplifier (HF2LI, Zurich Instruments) with a sampling frequency of 10 kHz (Fig. 1c). A good correlation between the voltage output of the sensor and the acoustic pressure is shown. The sensitivity *S* of the piezoelectric sensor is typically defined as $S = \delta((U-U_0)/U_0)/\delta p$ *(33)*, where $U_0 = 0.1$ mV is the initial voltage of the sensor without pressure, and *U* is the voltage response under the applied pressure *p*. The peak sensitivity reaches $3.5 \times 10^6$ kPa$^{-1}$ (Fig. 1e), which is approximately one order of magnitude larger than the highest sensitivity of $3.8 \times 10^5$ kPa$^{-1}$ *(34)* obtained from the current state-of-the-art pressure sensors.

Importantly, at pressure of 1 mPa or even lower, the sensor still exhibits a clear decay in the voltage response, indicating that the sensor can effectively detect pressure variations of 1 mPa or less (Fig. 1d). This value is lower than the lowest pressure detection limit of recently reported pressure sensors, 2 mPa *(3)*. The sensor thereby demonstrates supreme pressure-sensing performance in terms of sensitivity and pressure detection limit, superior to the state-of-the-art pressure sensors reported so far (Fig. 1b). Moreover, based on the pressure detection limit of 1 mPa and the sensor size of ~5 × 5 mm$^2$, the sensor can detect tiny force as low as 25 nanonewtons.

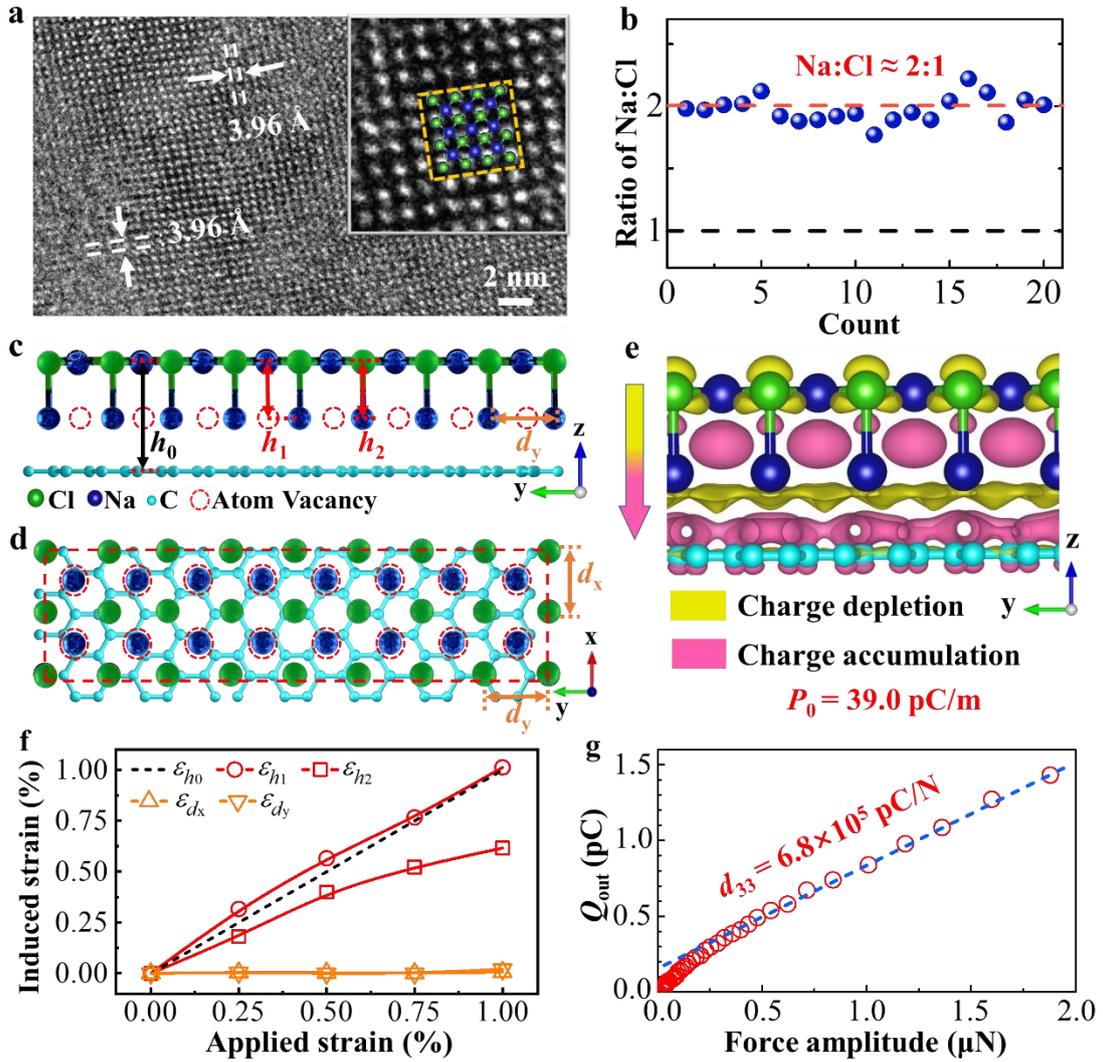

**Figure 2 | Mechanism of the piezoelectric sensor with Na$_2$Cl crystals. a,** High-resolution transmission electron microscopy (TEM) image of Na$_2$Cl crystals observed in the GO membrane. The inset shows a zoomed-in area of the high-resolution image and the Na$_2$Cl model, in which the Na and Cl atoms are shown as blue and green spheres respectively. **b,** Na:Cl ratio for the Na-Cl@GO membrane (from 20 different regions in the dark-field TEM images). Side view (**c**) and top view (**d**) of the 2D Na$_2$Cl crystal structure on the graphene base (Na$_2$Cl@graphene). Here, the graphene base represents the part of GO without functional groups. $h_1$ is the average displacement between the Na atom in the upper layer of the Na$_2$Cl crystal and the atom vacancies in the bottom layer of the Na$_2$Cl crystal; $h_2$ is the average displacement between the Cl atom in the upper layer of the Na$_2$Cl crystal and the Na atom in the bottom layer of the Na$_2$Cl crystal; $d_x$ and $d_y$ are the distances between adjacent atoms in each layer in x and y direction. **e,** Differential charge density (DCD) of Na$_2$Cl@graphene. **f,** Relative changes in different distances (induced strain) as a function of applied strain. **g,** Piezoelectric constant obtained from the decay process of the acoustic pressure. The dashed blue line is the tangent line on the curve in the figure and the slope of tangent line represents the piezoelectric coefficient ($d_{33}$).

We attribute the observed piezoelectric behaviors to the 2D $Na_2Cl$ crystals within the GO layers in the sensor. This $Na_2Cl$ crystal was discovered in our previous work (*35*), originating from the ion–π interactions (*36*) between the ions and the π-conjugated system in the graphitic surface (*37, 38*). Different from the ordinary NaCl crystal, the 2D $Na_2Cl$ crystal has a Na:Cl atom ratio of 2:1 and contains atom vacancies where Cl atoms should stay (Fig. 2c); here, we refer to these anion vacancies as "ghost atoms". These Cl atom vacancies in the crystal lead to very high asymmetry, while their periodic arrangement in the crystal guarantees stability.

The existence of 2D $Na_2Cl$ crystals in the sensor was confirmed by transmission electron microscopy (TEM) experiments. The high-resolution TEM image shows a square structure with a lattice spacing of 3.96 ± 0.03 Å (Fig. 2a). The enlarged image overlays well with the $Na_2Cl$ structure in our previous work (*35*), as shown by the inset in Fig. 2a, in which the Na and Cl atoms are shown as blue and green spheres, respectively. The dark-field TEM images demonstrate that the Na:Cl ratio is approximately 2:1 (Fig. 2b and Fig. S4).

We then performed density functional theory (DFT) computations to illustrate the underlying physics. The structure of the 2D $Na_2Cl$ crystal based on the graphene surface ($Na_2Cl$@graphene) is shown in Fig. 2c and 2d. Here, the graphene surface is considered as the part of GO membrane without functional groups. The electric polarization direction is shown in the line profile of the interfacial differential charge densities (DCDs) in the out-of-plane direction (Fig. 2e). By analyzing the Bader charges of the atoms, a very large spontaneous out-of-plane polarization *($P_0$)*, serving as one of crucial parameters for piezoelectric materials (*16, 39*), was obtained for the $Na_2Cl$@graphene (Supplementary Information, Materials and Methods), namely, $P_0$ = 39.0 pC/m. This value is nearly two orders of magnitude larger than the out-of-plane polarization of 0.6 pC/m of $MoS_2$/$WS_2$ (*17*).

The induced mechanical deformation of $Na_2Cl$@graphene under external pressure was then investigated by varying the distance between the Na atom in the upper layer of the $Na_2Cl$ crystal and the graphene surface using DFT simulations (Supplementary Information, Materials and Methods). Fig. 2f clearly shows that in the out-of-plane direction, the average displacement ($h_1$) between the Na atom in the upper layer of the $Na_2Cl$ crystal and the atom vacancies in the bottom layer of the $Na_2Cl$ crystal is much larger than the average displacement ($h_2$) between the Cl atom in the upper layer of the $Na_2Cl$ crystal and the Na atom in the bottom layer of the $Na_2Cl$ crystal under stress. In contrast, along the in-plane direction, the distances between adjacent atoms ($d_x$ and $d_y$) in each layer remain approximately constant. This indicates that the atom vacancies make the $Na_2Cl$ crystal susceptible to deformation along the direction of applied stress, while maintaining structural stability.

The very large intrinsic polarization suggests an extraordinarily large piezoelectric coefficient in the sensor (*40*). According to the stress-output charge relation from the acoustic excitation of the tuning fork (Fig. 2g), we experimentally obtained a piezoelectric coefficient ($d_{33}$) up to $6.8 \times 10^5$ pC/N in the range of 0.025–2 μN (corresponding to the pressure range of 1–80 mPa). This value is larger than the $d_{33}$ of $\sim 2 \times 10^5$ pC/N in cubic fluorite gadolinium-doped $CeO_{2-x}$ films obtained by rearranging oxygen vacancies under very large electric fields at millihertz frequencies (*18*).

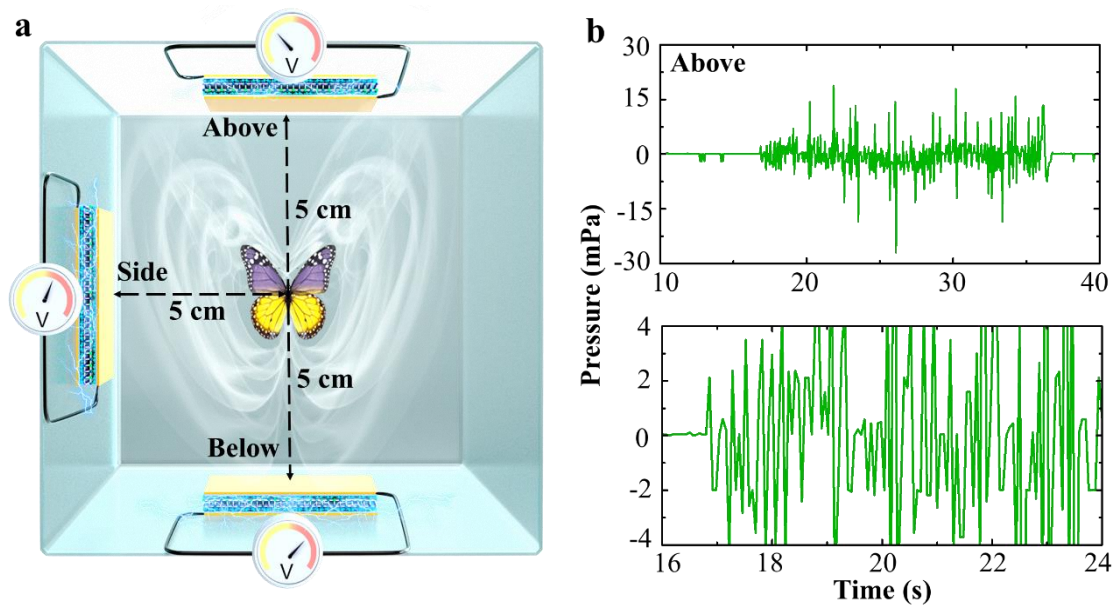

**Figure 3 | Detection of tiny airflow fluctuations surrounding a flapping butterfly with the piezoelectric sensor. a,** A schematic of the real-time monitoring of airflow pressure surrounding a flapping butterfly. Three piezoelectric sensors are at a 5 cm distance from a flapping butterfly. **b,** Air pressure variations in real time. The top picture depicts air pressure variations obtained from the sensor above a flapping butterfly, and the bottom picture is a partially enlarged view of the top picture.

The famous butterfly effect, a metaphor for chaos theory that posits "Does the flap of a butterfly's wings in Brazil set off a tornado in Texas?" (*41*), reflects sensitive dependence on an initial condition, with extremely small signals leading to extraordinarily large responses. To our knowledge, the airflow variation caused by a flapping butterfly has not been well measured because the variation is quite small (*42*). Using the presented sensor, the open-circuit voltage response to the airflow variation at a 5 cm distance from a flapping butterfly was obtained, and the movement of the butterfly and the real-time motoring of voltage response were shown in the movies (Supplementary Text, section 9). The corresponding pressure variations above, on the side of, and below the butterfly were ~3.9 mPa, ~6.8 mPa, and ~14.7 mPa, respectively (Fig. 3b; Fig. S12b), calculated by the conversion relation between the pressure and voltage response (Supplementary Text, section 7). The pressure variations of 2 mPa or even lower can be clearly distinguished by the sensor (Fig. 3b). These variations in the airflow pressures are extremely small, only approximately one millionth of the airflow pressure caused by a tornado. A wing-beat frequency of ~5 Hz was obtained (Supplementary Text, Fig. S12c), which is consistent with the results from other studies on flapping frequency (*42*).

The sensor also exhibits a very high sensitivity in the high-pressure range. We used a force gauge (Mark-10, FS05) to produce a series of mechanical loads in the pressure range of 0.1 to 100 kPa. The voltage response of the sensor was measured using an electrochemical workstation (CHI760E) (Fig. S6a). The output voltage shows a cyclic and step-like response, which increases from ~0.06 V to ~0.80 V and is well correlated with the applied mechanical load (Fig. S6a). The sensitivity of the sensor was then calculated. The sensitivity decreases from $2.0 \times 10^4$ kPa$^{-1}$ to $1.2 \times 10^2$ kPa$^{-1}$ when the applied pressure is varied in the range of 0.1–100 kPa (Fig. S6b).

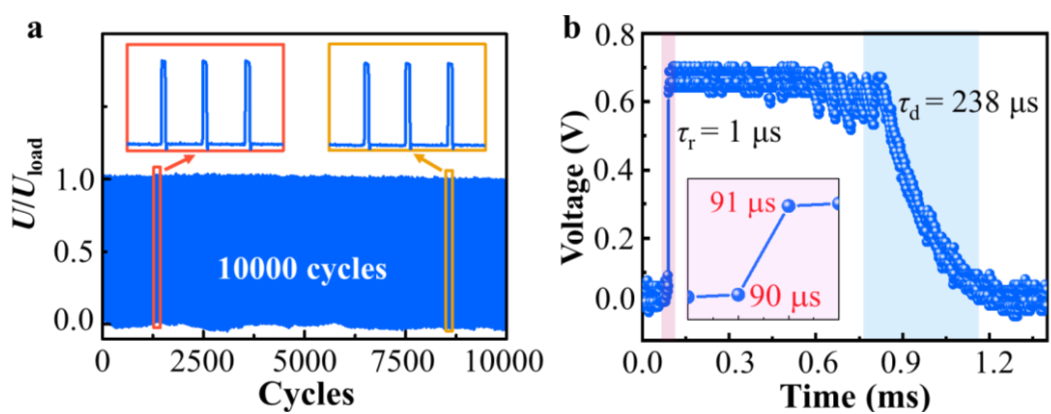

**Figure 4 | Mechanical stability and response time of the piezoelectric sensor. a**, Relative change in voltage ($U$-$U_{load}$) during cyclic loading/unloading experiments under the loaded pressure of 80 kPa. $U_{load}$ = 0.431 V is the average voltage when the sensor is loaded at the initial stage. The inserts are two partially enlarged views. **b**, Response and recovery times of the sensor under loaded pressure of 98 kPa measured by oscilloscope with a time step of 1 μs. The light red-shaded and light blue-shaded areas show the response and recovery processes of the sensor. The inset is a partially enlarged view.

In addition, the sensor has excellent mechanical stability and a fast response. The peak voltage due to a mechanical pressure of 80 kPa decreases by less than 3% when the sensor is subjected to over 10000 cycles of loading and unloading (Fig. 4a). A response time ($\tau_r$) of 1 μs and recovery time ($\tau_d$) of 238 μs were obtained with an oscilloscope (Tektronix, MSO44 4-BW-200) (Fig. 4b). The detected response time of 1 μs is actually the limit of our measurement set-up instead of the intrinsic limit of the sensor. The excellent mechanical stability and fast response ensure its application in the area of sensitive pressure sensing.

In summary, we developed an ultrasensitive piezoelectric sensor made of multilayered Na$_2$Cl@GO membrane prepared by spray-coating method. The sensor exhibits excellent pressure-sensing performance, including ultrahigh sensitivity and an extremely low detection pressure limit under ambient conditions. DFT computations reveal that the ultrasensitive piezoelectric response of the sensor can be attributed to the unique atom vacancies of the Cl atoms ("ghost atoms") in $Na_2Cl$ crystal. The high percentage of atom vacancies induces a very large intrinsic polarization and thus the sensor exhibits an extraordinarily large piezoelectric constant and sensitive response to mechanical pressure, while the periodic arrangement of the atom vacancies guarantees structural stability.

Notably, the sensor can detect nanonewton forces as low as 25 nanonewtons, based on its low pressure detection limit of ~1 mPa and device size of ~5 × 5 $mm^2$. The extremely small airflow fluctuations at a 5 cm distance from a flapping butterfly can be clearly detected, which have long been the elusive tiny signals in the famous "butterfly effect". Moreover, the sensitivity and the pressure detection limit can be further optimized by increasing the content of the $Na_2Cl$ crystals within GO membrane. Similarly, the size of the sensor can be even smaller while keeping the excellent pressure-sensing performance by further increasing the crystal content. Considering its excellent mechanical stability, fast response and self-powered capability, the sensor shows great potential in applications that require the high sensitivity to external mechanical stimuli and the detection of the tiny forces with only several nanonewtons, even for monitoring the very fast-changed signals of systems in very small spaces such as the brain and other organs. Finally, the presented sensor can serve as a seminal sensor to detect other tiny signals, including electrical, magnetic, optical, and thermal signals, by converting these signals to mechanical forces, and thus the applications can be further greatly extended; for example, combining the sensor with a magnet, the very small magnetic fields can be measured by detecting the tiny force acting on the magnet.